\documentclass[aps,prb,twocolumn,superscriptaddress,showpacs]{revtex4}
\usepackage{graphicx}
\usepackage{mathrsfs}
\usepackage{bm}

\begin{document}
\title{Microscopic theory of quantum anomalous Hall effect in graphene}
\author{Zhenhua Qiao}
\affiliation{Department of Physics, The University of Texas at
Austin, Austin, Texas 78712, USA}
\author{Hua Jiang}
\affiliation{International Center for Quantum Materials, Peking University, Beijing 100871, China}
\author{Xiao Li}
\affiliation{Department of Physics, The University of Texas at
Austin, Austin, Texas 78712, USA}
\author{Yugui Yao}
\affiliation{School of Physics, Beijing Institute of Technology, Beijing 100081, China}
\author{Qian Niu}
\affiliation{Department of Physics, The University of Texas at
Austin, Austin, Texas 78712, USA}\affiliation{International Center for Quantum Materials, Peking University, Beijing 100871, China}

\begin{abstract}
We present a microscopic theory to give a physical picture of the formation of quantum anomalous Hall (QAH) effect in graphene due to a joint effect of Rashba spin-orbit coupling $\lambda_R$ and exchange field $M$. Based on a continuum model at valley $K$ or $K'$, we show that there exist two distinct physical origins of QAH effect at two different limits. For $M/\lambda_R\gg1$, the quantization of Hall conductance in the absence of Landau-level quantization can be regarded as a summation of the topological charges carried by Skyrmions from real spin textures and Merons from \emph{AB} sublattice pseudo-spin textures; while for $\lambda_R/M\gg1$, the four-band low-energy model Hamiltonian is reduced to a two-band extended Haldane's model, giving rise to a nonzero Chern number $\mathcal{C}=1$ at either $K$ or $K'$. In the presence of staggered \emph{AB} sublattice potential $U$, a topological phase transition occurs at $U=M$ from a QAH phase to a quantum valley-Hall phase. We further find that the band gap responses at $K$ and $K'$ are different when $\lambda_R$, $M$, and $U$ are simultaneously considered. We also show that the QAH phase is robust against weak intrinsic spin-orbit coupling $\lambda_{SO}$, and it transitions a trivial phase when $\lambda_{SO}>(\sqrt{M^2+\lambda^2_R}+M)/2$. Moreover, we use a tight-binding model to reproduce the ab-initio method obtained band structures through doping magnetic atoms on $3\times3$ and $4\times4$ supercells of graphene, and explain the physical mechanisms of opening a nontrivial bulk gap to realize the QAH effect in different supercells of graphene.
\end{abstract}

\maketitle

\section{Introduction}
In 1879, Edward H. Hall discovered that when an electric field flows through a conductor in the presence of a perpendicular magnetic field, charge carriers subjected to the Lorentz force are pushed to one side of the conductor. At equilibrium, the carrier accumulation generates a transverse bias to balance the Lorentz force. This is the famous ``Hall effect''. In two-dimensional electron systems, the quantized version of Hall Effect was observed due to Landau quantization,~\cite{QHE} which is characterized by a precisely quantized Hall conductance, i.e.
$\sigma_{xy}=\mathcal{C}~{e^2}/{h}$, where $\mathcal{C}$ is known as the TKNN number or Chern number.~\cite{Thouless,Kohmoto}

To produce a Hall effect, breaking time-reversal symmetry is an essential condition. In addition to magnetic field, an internal magnetization coupled with spin-orbit coupling could also give rise to the Hall effect. To distinguish from the ordinary Hall effect, this magnetization-induced one was called ``anomalous'' Hall effect. Although it has been experimentally observed for over one century, the physical origin of anomalous Hall effect is still unclear. In general, the mechanism of anomalous Hall effect is classified as extrinsic or intrinsic according to its origins. The extrinsic one arises from the spin-dependent scattering impurities, while the latter one can be expressed in terms of Berry-phase curvatures in the crystal momentum space.~\cite{AHE-RMP,BerryPhase-Niu}

Similar to the quantization of the ordinary Hall effect, the anomalous Hall effect was also predicted to be quantized by Haldane in a honey-comb lattice toy model with vanishing magnetic field.~\cite{haldane} Subsequently, other proposals were made toward the realization of quantum anomalous Hall (QAH) effect, i.e. in Mercury-based quantum wells,~\cite{ChaoxingLiu} disorder-induced Anderson insulator,~\cite{Nagaosa} optical lattices,~\cite{CongjunWu} and magnetic topological insulators.~\cite{RuiYu} Despite the theoretical progress, the QAH effect has yet observed experimentally. In a recent paper,~\cite{qiao,Tse} we found that graphene shows great potential to host the long-sought QAH state in the presence of Rashba spin-orbit coupling and exchange field. Based on the state-of-the-art first-principles calculation method, researchers~\cite{Ding,HongbinZhang} further demonstrated that this QAH phase could be engineered via doping \emph{3d} or \emph{5d} transition metal atoms on the hollow sites of graphene. The final realization of QAH effect will not only enable the application of novel quantum devices due to the dissipationless nature, but also mark the ultimate achievement of a clear understanding of the intrinsic mechanism of the anomalous Hall effect.~\cite{AHE-RMP}

In this paper, we demonstrate a microscopic theory to study the physical origins of the QAH effect in graphene due to the presence of both Rashba spin-orbit coupling and exchange using a low-energy continuum model. In the limit of strong exchange field and weak Rashba spin-orbit coupling, the quantization of the Hall conductance should be attributed to the real spin texture-induced Skyrmions and \emph{AB} sublattice pseudo-spin texture-induced Merons. While in the other limit, i.e. weak exchange field and strong Rashba spin-orbit coupling, the four-band low-energy model can be reduced to a two-band extended Haldane's model. We also show that this QAH phase is robust against weak staggered \emph{AB} sublattice potentials or intrinsic spin-orbit coupling, which is present in real materials. Using a tight-binding method, we reproduce all the ab-initio obtained band structures of doping magnetic atoms in $3\times3$ or $4\times4$ supercells of graphene. And we give an explanation of the formation mechanism of the QAH effect in $3\times3$ or $4\times4$ supercell of graphene.

The remainder of the paper is organized as follows. In
Sec.~{\ref{section1}}, we present tight-binding and continuum models of graphene in the presence of Rashba spin-orbit coupling, intrinsic spin-orbit coupling, exchange field, and staggered AB sublattice potential. Section~{\ref{section2}} discusses the physical origin of the
quantum anomalous Hall effect in graphene at two different limits using a continuum model. In Sec.~{\ref{section3}}, we show the robustness of the quantum anomalous Hall state in the presence of either staggered \emph{AB} sublattice potential or intrinsic spin-orbit coupling. In Sec.~{\ref{section4}}, we use a tight-binding model to explain a recent ab-initio work about realizing the quantum anomalous Hall effect in $3\times3$ and $4\times4$ supercell of graphene. A brief summary is given in Sec.~{\ref{section5}} to close the paper.

\section{Model Hamiltonian of Graphene} \label{section1}
The real space $\pi$-orbital tight-binding Hamiltonian of single layer graphene in the presence of Rashba/intrisic spin-orbit coupling, exchange field and staggered AB sublattice potentials is written as~\cite{Kane,Sheng,qiao}:
\begin{eqnarray}
H({\bm r})=H_0({\bm r})+H_R({\bm r})+H_{SO}({\bm r})+H_M({\bm r})+H_U({\bm r}),
\end{eqnarray}
where each term is given by
\begin{eqnarray}
&&H_0({\bm r})=- t \sum_{\langle{ij}\rangle; \alpha }{ c^\dagger_{i
\alpha}c_{j\alpha}}; \nonumber \\
&&H_R({\bm r})={i} t_{R}\sum_{\langle{ij}\rangle; \alpha, \beta
}\hat{\mathbf{e}}_{z}{\cdot}({{\bm s}_{\alpha
\beta}}{\times} {\mathbf{d}}_{ij})c^\dagger_{i \alpha } c_{j \beta };  \nonumber \\
&&H_{SO}({\bm r})=\frac{2i}{\sqrt{3}}t_{SO}\sum_{\ll{ij}\gg}
{c^\dagger_{i}{\bm s}{\cdot}(\mathbf{d}_{kj}{\times}\mathbf{d}_{ik})c_{j}}; \nonumber \\
&&H_M({\bm r})=M \sum_{i;\alpha,\beta}{ c^\dagger_{i\alpha}{\bm s}^z_{\alpha\beta}c_{i\beta}}; \nonumber \\
&&H_U({\bm r})=\sum_{i;\alpha}{ c^\dagger_{i\alpha} V_{i} c_{i\alpha}}.\label{SGHamiltonian} \nonumber
\end{eqnarray}
Here, $c^\dag_{i\alpha}$ and $c_{i\alpha}$ are $\pi$-orbital creation and annihilation operators for an electron with spin $\alpha$ on site $i$. The first term $H_0$ represents the nearest neighbor hopping with amplitude  $t=~2.6~eV$. The second term $H_R$ describes the Rashba spin-orbit coupling with ${\mathbf{d}}_{ij}$ being a lattice vector pointing from site $j$ to site $i$. The third term $H_{SO}$ is the intrinsic spin-orbit coupling with $k$ connecting the next-nearest neighbor sites $i$ and $j$. $\langle\rangle$/$\ll\gg$ runs over all the nearest/next-nearest neighbor hopping sites. The fourth term and the last term correspond to the exchange field and staggered \emph{AB}-sublattice potentials, respectively. We set $V_i=+U$ at \emph{A}-type sublattices and $V_i=-U$ at \emph{B}-type sublattices. $\alpha$ and $\beta$ denote spin indices, and ${\bm s}$ are the spin Pauli matrices.

By performing a Fourier transformation, the real space Hamiltonian in Eq.~(\ref{SGHamiltonian}) is converted to a $4\times 4$ matrix $H({\bm k})$ in the momentum space. In this paper, we choose the lattice unit vectors to be
\begin{eqnarray}
{\bm a}_1=\frac{a}{2}(2\sqrt{3},~0), {\bm a}_2=\frac{a}{2}(\sqrt{3},~3),
\end{eqnarray}
and the corresponding reciprocal-lattice vectors are given by
\begin{eqnarray}
{\bm b}_1=\frac{2\pi}{a}(\frac{1}{\sqrt{3}},~\frac{-1}{3}), {\bm b}_2=\frac{2\pi}{a}(0,~\frac{2}{3}).
\end{eqnarray}
where $a=1.42~{\AA}$ is the distance between nearest neighbor carbon-carbon atoms, and we set $a$ to be unity in the following calculation for simplicity.
On the basis of \{$\psi_{A\uparrow}$, $\psi_{A\downarrow}$, $\psi_{B\uparrow}$, $\psi_{B\downarrow}$\}, the corresponding momentum-space Hamiltonian of each term is listed in the following.

(A) Nearest-neighbor hopping term:
\begin{eqnarray}
H_{0}(\bm k)=-t \left[
\begin{array}{cccc}
0 & \gamma_0 \\
{\gamma^*_0} & 0
\end{array}
\right],\label{H0-k}
\end{eqnarray}
with
\begin{eqnarray}
\gamma_0&=&[(2 \cos \frac{\sqrt{3} k_x}{2}  \cos \frac{k_y}{2} + \cos {k_y})  \nonumber \\
&& +i (2 \cos \frac{\sqrt{3} k_x}{2}  \sin \frac{k_y}{2} - \sin {k_y})] {\bm{1}}_{s},\nonumber
\end{eqnarray}
where ${\bm{1}}_{s}$ is a $2\times2$ identity matrix.

(B) Rashba spin-orbit coupling term:
\begin{eqnarray}
H_{R}(\bm k)=t_R \left[
\begin{array}{cccc}
0 & \gamma_R \\
{\gamma^*_R} & 0
\end{array}
\right],
\end{eqnarray}
with
\begin{eqnarray}
\gamma_R&=&[(\cos \frac{\sqrt{3}k_x}{2}  \sin \frac{k_y}{2}+\sin {k_y}) \nonumber \\
&&-i(\cos \frac{\sqrt{3}k_x}{2}  \cos \frac{k_y}{2}-\cos {k_y})]{\bm s}_x  \nonumber \\
&&-\sqrt{3}\sin \frac{\sqrt{3}k_x}{2}  (i \sin \frac{k_y}{2}+\cos \frac{k_y}{2}){\bm s}_y.\nonumber
\end{eqnarray}

(C) Intrinsic spin-orbit coupling term:
\begin{eqnarray}
H_{SO}(\bm k)=t_{SO} \left[
\begin{array}{cccc}
\gamma_{SO} & 0 \\
0 & -\gamma_{SO}
\end{array}
\right],
\end{eqnarray}
where
\begin{eqnarray}
\gamma_{SO}=-4t_{SO}~\sin \frac{\sqrt{3}k_x}{2} (\cos \frac{3k_y}{2}-\cos \frac{\sqrt{3}k_x}{2}) {\bm s}_z.\nonumber
\end{eqnarray}

(D) Exchange field term:
\begin{eqnarray}
H_M(\bm k)=M \left[
\begin{array}{cccc}
{\bm s}_z & 0 \\
0 & {\bm s}_z
\end{array}
\right].
\end{eqnarray}

(E) Staggered AB sublattice potential term:
\begin{eqnarray}
H_U(\bm k)=U \left[
\begin{array}{cccc}
{\bm{1}}_{s} & 0 \\
0 & -{\bm{1}}_{s}
\end{array}
\right].\label{HV-k}
\end{eqnarray}

By directly diagonalizing $H({\bm k})$ at each crystal momentum $\bm k$, one can easily obtain the bulk band structure. As reported in a recent paper~\cite{qiao}, we found that a nontrivial bulk gap opens when both the Rashba spin-orbit coupling $t_R$ and exchange field $M$ are considered simultaneously. Through calculating the Chern number by integrating the Berry curvatures in the first Brillouin zone, we found that the resulting Chern number is nonzero, indicating a quantum anomalous Hall state. The central issue in this paper is to give a physical picture to understand the formation of this nontrivial state. Therefore, the study the low-energy effective model is required.

Through expanding the tight-binding Hamiltonian in Eq.(\ref{H0-k}) - Eq.(\ref{HV-k}) at the vicinity of valleys $K$ and $K'$, i.e. ($k_x, ~ k_y$)=($\pm 4 \pi/3{\sqrt{3}},~0$), the low-energy effective model Hamiltonian of each term at valleys $K$ and $K'$ is summarized as following on the basis of \{$\psi_{A\uparrow}$, $\psi_{A\downarrow}$, $\psi_{B\uparrow}$, $\psi_{B\downarrow}$\}:
\begin{eqnarray}
&&h_0(\bm{k})=v(\eta \sigma_x k_x + \sigma_y k_y )\bm{1}_s; \\
&&h_R(\bm{k})=\frac{\lambda_{\mathrm{R}}}{2}(\eta\sigma_x s_y - \sigma_y s_x); \\
&&h_{SO}(\bm{k})=\eta \lambda_{SO}\sigma_z {s}_z ; \\
&&h_M(\bm{k})=M\bm{1}_\sigma {s}_z;  \\
&&h_U(\bm{k})=U\sigma_z \bm{1}_{s}.
\end{eqnarray}
Here, $\eta =\pm 1$ labels valley degrees of freedom; $\bm{\sigma}$ are Pauli matrices representing the \emph{AB}-sublattice pseudo-spin degrees of freedom. The Fermi velocity, Rashba spin-orbit coupling, and intrinsic spin-orbit coupling are given by $v = 3t/2$, $\lambda_{R} = 3t_{R}$, and $\lambda_{SO} = 3\sqrt{3} t_{SO}$, respectively.

\section{Physical Origin of Quantum Anomalous Hall Effect}\label{section2}
When the Rashba spin-orbit coupling $\lambda_{R}$ and exchange field $M$ are taken into account simultaneously, the continuum model Hamiltonian is
\begin{eqnarray}
H({\bm k})=h_0({\bm k})+h_R({\bm k})+h_M({\bm k})\label{Ham-QAH}.
\end{eqnarray}
In Refs.~[\onlinecite{qiao},\onlinecite{Tse}], we have pointed out that a nontrivial bulk band gap opens up as long as $\lambda_R$ and $M$ are nonzero. Based on the Kubo formula, when the Fermi energy lies within the bulk band gap, the corresponding Hall conductance $\sigma_{xy}$ is shown to be quantized as:
\begin{eqnarray}
\sigma_{xy}&=&\mathcal{C}\frac{e^2}{h} {\rm sgn}(M),
\end{eqnarray}
where the Chern number is $\mathcal{C}=2$ and can be calculated from
\begin{eqnarray}
\mathcal{C}=\frac{1}{2\pi}\sum_{K,K'}\sum_{n=1,2}\int^{+\infty}_{-\infty} d k_x d k_y \Omega _n(k_x,k_y).
\end{eqnarray}
The sum is taken over both valley $K/K'$ and $n$ occupied valence bands below the bulk band gap. $\Omega _n$ is the momentum-space Berry curvature of the $n$-th band, and can be obtained through the following formula
\begin{eqnarray}
\Omega_n(\bm{k})=-{\sum_{n^{\prime} \neq n}} {\frac{2 {\rm {Im}}
\langle \psi_{n \bm{k}}|v_x|\psi_{n^\prime \bm{k}} \rangle \langle
\psi_{n^\prime \bm{k}}|v_y|\psi_{n \bm{k}} \rangle }
{(\omega_{n^\prime}-\omega_{n})^2}}, \label{berry}
\end{eqnarray}
where $\omega_n\equiv E_n/ \hbar$, and $v_{x(y)}$ is the Fermi velocity
operator.

\begin{figure}
\includegraphics[width=7cm,angle=0]{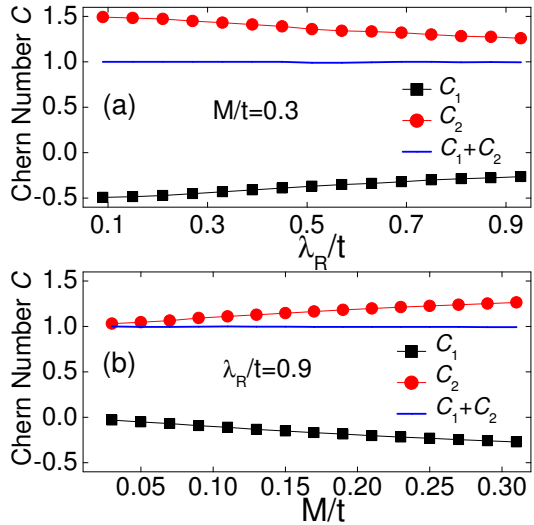}
\caption{(Color online) Upper panel: Chern number of each valence band as a function of Rashba spin-orbit coupling $\lambda_R$ at fixed exchange field $M/t=0.3$; Lower panel: Chern number of each valence band as a function of exchange field M at fixed Rashba spin-orbit coupling $\lambda_R/t=0.9$. The solid (blue) line represents the summation of the two valence bands. The cutoff of $k_x$ and $k_y$ are set to be $k_0=\pi/2a$.} \label{ChernNumberDependence}
\end{figure}

It is already known that the Chern numbers of valleys $K$ and $K'$ are equal,~\cite{Tse} i.e. $\mathcal{C}_K=\mathcal{C}_{K'}=1$. However, the component from each valence band is still unclear. For clarity, we label the lowest valence band as the $1{st}$ band, and the other one close to the bulk band gap as the $2{nd}$ band. Figure~\ref{ChernNumberDependence} plots the Chern number of each band and the total Chern number as functions of exchange field $M$ and Rashba spin-orbit coupling $\lambda_R$. In Fig.~\ref{ChernNumberDependence}(a), the exchange field is fixed at $M/t=0.3$. We find that, for extremely small Rashba spin-orbit coupling $\lambda_R\rightarrow0$, the Chern number of the $1st$ valence band is half-quantized with a negative sign, i.e. $\mathcal{C}_1=-0.5$; and that of the $2nd$ valence band is one and half quantized, i.e. $\mathcal{C}_2=1.5$. When Rashba spin-orbit coupling $\lambda_R$ gradually increases, the absolute values of $\mathcal{C}_1$ and $\mathcal{C}_2$ are both reduced. On the contrary, in Fig.~\ref{ChernNumberDependence}(b) Rashba spin-orbit coupling is fixed to be $\lambda_R/t=0.3$. One observes that for extremely weak exchange field $M\rightarrow0$, the $1st$ valence band gives no contribution to the total Chern number, i.e. $\mathcal{C}_1=0$ and $\mathcal{C}_2=1$. Along with the increasing of the exchange field, $\mathcal{C}_{1}$ increases with a negative sign, while $\mathcal{C}_{2}$ is also linearly increased, keeping the total Chern number quantized to be $\mathcal{C}=1$.

Based on the analysis of the Chern number response at the two different limits, i.e. $M/\lambda_R \gg1$ and $\lambda_R/M \gg1$, one can imagine that the resulting Hall conductance quantization should correspond to different formation mechanisms. In the following, we give a clear understanding of the physical origins of the quantum anomalous Hall effect at the two different limits.

\subsection{$M/\lambda_R \gg 1$ Limit: Skyrmion and Meron}
From Fig.~\ref{ChernNumberDependence}(a), one can note that both valence bands
contribute to the total Hall conductance, therefore the four-band Hamiltonian can not be reduced
to a two-band effective Hamiltonian model. In the following, we study the origin of the Hall conductance from each band. In our studied single layer graphene system, there are two
kinds of spin degrees of freedom: real spin $\bm{s}$ and \emph{AB} sublattice
pseudo-spin $\bm {\sigma}$. On the basis of \{$\psi_{A\uparrow}$, $\psi_{A\downarrow}$, $\psi_{B\uparrow}$, $\psi_{B\downarrow}$\}, the real spin and pseudo-spin components can be evaluated through
\begin{eqnarray}
\langle {\bm s}_i \rangle &=& \langle \psi|~ {\bm 1}_{\sigma} \otimes {\bm s}_i ~|\psi \rangle,
\nonumber \\
\langle {\bm \sigma}_i \rangle &= &\langle \psi|~  {\bm \sigma}_i  \otimes {\bm 1}_{s}~
|\psi \rangle,
\end{eqnarray}
where $i=\{x,~y,~z\}$, and $|\psi \rangle$ is a $4\times 1$ eigenvector.

\begin{figure*}
\includegraphics[width=15cm,angle=0]{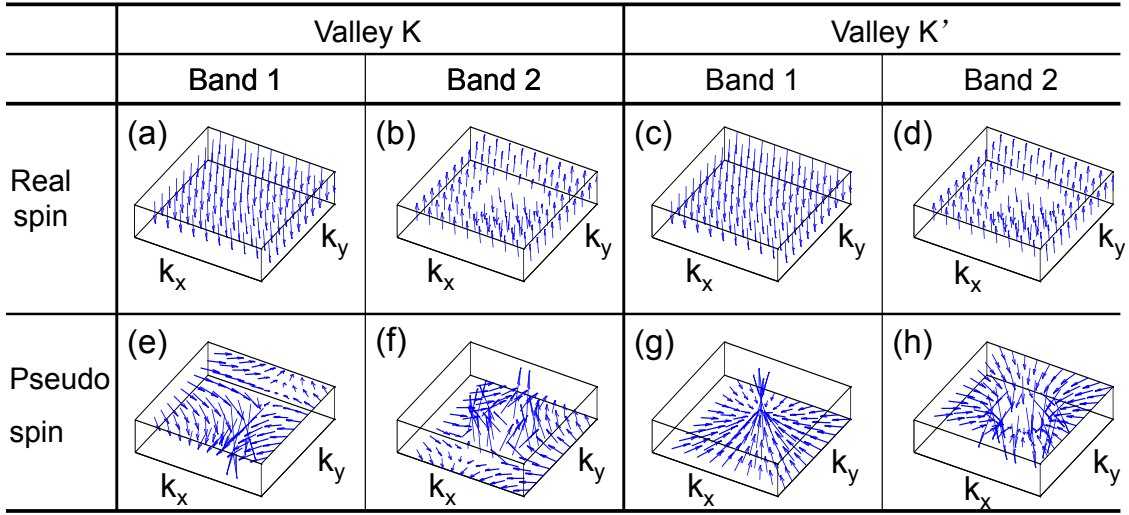}
\caption{Real spin and pseudo-spin textures of the two valence bands of graphene below the band gap at valleys $K$ and $K'$ in the limit of $M/\lambda_R\gg1$ (here, we set $M=0.8$ and $\lambda_R=0.12$). (a)-(d): Real spin textures of the $1st$ and $2nd$ valence bands at valleys $K$ and $K'$. (a) and (c): Spins point toward the south pole uniformly; (b) and (d): Spins at the center point toward the south pole while those far away from the center point toward the north pole, which indicates a Skyrmion. (e)-(h): Pseudo-spin textures of the $1st$ and $2nd$ valence bands at valleys $K$ and $K'$. (e) and (f): In-plane pseudo-spin components share the similar winding pattern pointing toward the center, but out-of-plane pseudo-spin components point toward south and north poles, respectively; (g)-(h): In-plane pseudo-spin components point toward the center without any winding, while out-of-plane pseudo-spin components point toward north and south poles, respectively. Each pseudo spin texture corresponds to a Meron or half-Skyrmion.} \label{SpinTexture}
\end{figure*}

In Fig.~\ref{SpinTexture}, we exhibit the spin textures of the real spin and $AB$ sublattice pseudo-spin at valleys $K$ and $K'$ for the two valence bands below the bulk band gap, respectively. One can observe that the real spin textures of the two valence bands of valley $K$ are exactly the same as that of valley $K'$ [see Fig.~\ref{SpinTexture}(a)-Fig.~\ref{SpinTexture}(d)], \emph{i.e.} for the $1st$ band the spins are uniformly pointing toward the south pole in the whole momentum space, contributing nothing to the total Chern number; while for the $2nd$ band the spins close to valleys $K$ and $K'$ point toward the south pole whereas those far away from the center point toward the opposite north pole [Note that there should exist a circular region with spin lying within the in-plane of the equator], which phenomenally corresponds to a Skyrmion that contributes to one topological charge.\cite{Skyrmion}

However, it becomes more complicated for the $AB$ sublattice pseudo-spin textures [see Fig.~\ref{SpinTexture}(e)-Fig.~\ref{SpinTexture}(h)]. For valley $K$, one can see that the in-plane pseudo-spin components of both valence bands point toward the center with the same winding pattern; the out-of-plane pseudo-spin components close to the center point toward either the south pole ($1st$ band) or north pole ($2nd$ band), but those far away from the center are vanishing. For valley $K'$, one finds that the pseudo spin textures are distinct from those at valley $K$. For example, the in-plane pseudo spins point toward the center without any winding, while the out-of-plane pseudo spins only exist near the valleys and point to either the north pole ($1st$ band) or south pole ($2nd$ band). All these suggest that each of the four different pseudo spin textures corresponds to a Meron, i.e. half-Skyrmion.\cite{HalfSkyrmion}

To confirm the above analysis, we should precisely calculate the Chern number (or topological charge) resulting from each special real spin or pseudo-spin texture using the following formula
\begin{eqnarray}
n=\frac{1}{4\pi}\int\int{dk_x dk_y (\partial_{k_x} \hat{\bm h} \times \partial_{k_y} \hat{\bm h})\cdot \hat{\bm h}},\label{TopologicalCharge}
\end{eqnarray}
where $n$ is a topological charge counting the number of times a unit vector $\hat{\bm h}(k)$ winding around the unit sphere as a function of $\bm{k}$. $\hat{\bm h}(\bm {k}) \equiv {\bm h}({\bm k})/|{\bm h}({\bm k})|$ with ${\bm h}({\bm k})$ representing the projection of the Hamiltonian shown in Eq.~(\ref{Ham-QAH}) into the real spin or pseudo-spin space. For $M/{\lambda_R}\gg1$, our numerical calculation shows that
\begin{eqnarray}
&&n^{K}_{1s}=n^{K'}_{1s}=0; \\
&&n^{K}_{2s}=n^{K'}_{2s}\simeq1.0; \\
&&n^{K}_{1\sigma}=n^{K'}_{1\sigma}\simeq-0.5;\label{sigma1} \\
&&n^{K}_{2\sigma}=n^{K'}_{2\sigma}\simeq 0.5 \label{sigma2}.
\end{eqnarray}
Therefore the corresponding Chern numbers become
\begin{eqnarray}
&&\mathcal{C}^{\rm K}_{1}=n^{K}_{1s}+n^{K}_{1\sigma}=-0.5; \\
&&\mathcal{C}^{\rm K'}_{1}=n^{K'}_{1s}+n^{K'}_{1\sigma}=-0.5; \\
&&\mathcal{C}^{\rm K}_{2}=n^{K}_{2s}+n^{K}_{2\sigma}=1.5; \\
&&\mathcal{C}^{\rm K'}_{2}=n^{K'}_{2s}+n^{K'}_{2\sigma}=1.5; \\
&&\mathcal{C}^{\rm K}=\mathcal{C}^{\rm K}_{1}+\mathcal{C}^{\rm K}_{2}=1; \\
&&\mathcal{C}^{\rm K'}=\mathcal{C}^{\rm K'}_{1}+\mathcal{C}^{\rm K'}_{2}=1.
\end{eqnarray}
From the relationship shown in Eqs.~(\ref{sigma1}) and (\ref{sigma2}), one can find that though the spins point to opposite poles (for example, see the pseudo spin textures of the $1st$ valence band at $K$ and $K'$), their different winding patterns give rise to the same winding number.

Therefore, in the limit of $M/{\lambda_R} \gg 1$, the formation of the quantum anomalous Hall state originates from both Skyrmions carried by the real spin textures and Merons carried by the \emph{AB} sublattice pseudo-spin textures. Quantitatively, the pseudo-spin induced topological charges $n$ from $1st$ and $2nd$ valence are exactly opposite to cancel each other, which makes the real spin-induced Skyrmions from $2nd$ become the only source to achieve the quantized Hall conductance without external magnetic field.

\subsection{$\lambda_R/M \gg 1$ Limit: An Extended Haldane Model}
In the limit of strong Rashba spin-orbit coupling $\lambda_R$ and weak exchange field $M$, the total Chern number mainly comes from the $2nd$ valence band while the contribution from $1st$ valence band is negligible as shown in Fig.~\ref{ChernNumberDependence}(b), i.e. $\mathcal{C}_1\simeq0$ and $\mathcal{C}_2\simeq1$. This indicates that it is possible to obtain a reduced effective two-band model Hamiltonian through disregarding the high-energy bands. By reconstructing the basis to be \{$\psi_{A\uparrow}$, $\psi_{B\downarrow}$, $\psi_{B\uparrow}$, $\psi_{A\downarrow}$\}, the continuum Hamiltonian at valley $K$ is written as
\begin{eqnarray}
H_{\rm K}&=&\left[
\begin{array}{cccc}
M & 0 & v k_{-} & 0 \\ \nonumber 0 & -M & 0 &
v k_{+} \\ \nonumber v k_{+} & 0 & M & -i \lambda_R \\
\nonumber 0 & v k_{-} & i \lambda_R & -M
\end{array}
\right] =\left[
\begin{array}{cccc}
H_1 & T \\
T^{\dag} & H_2
\end{array}
\right]
\end{eqnarray}
where $H_1$ and $H_2$ represent the two block Hamiltonians on the basis
of ($\psi_{A\uparrow}$, $\psi_{B\downarrow}$) and ($\psi_{B\uparrow}$, $\psi_{A\downarrow}$),
respectively; $T$ couples the two different blocks.
At the vicinity of $K$, $H_1$ and $H_2$ correspond to the low energy band (\emph{i.e.} $\varepsilon=\pm M$) and high
energy band (\emph{i.e.} $\varepsilon=\pm\sqrt{M^2+\lambda^2_R}$), and
the coupling $T$ becomes extremely weak. Therefore, an effective
Hamiltonian can be obtained to describe the low-energy physics at $K$ point:
\begin{eqnarray}
&&H^{\rm K}_{\rm eff}\simeq H_{1}-T H^{-1}_{2} T^{\dag} = d_z \sigma_z+d_y\sigma_y+d_x \sigma_x, \label{effect-K1} \\ \nonumber
&&d_z=M(1-\frac{v^2}{\lambda^2_R}k^2); \\ \nonumber
&&d_y=-\frac{v^2}{\lambda_R}(k^2_x-k^2_y); \\ \nonumber
&&d_x=2\frac{v^2}{\lambda_R}k_x k_y.\\ \nonumber
\end{eqnarray}
Similarly, after reconstructing the basis to be \{$\psi_{B\uparrow}$, $\psi_{A\downarrow}$, $\psi_{A\uparrow}$, $\psi_{B\downarrow}$\}, the continuum Hamiltonian at at valley $K'$ becomes
\begin{eqnarray}
H_{\rm K'}=\left[
\begin{array}{cccc}
M & 0 & -v k_{-} & 0 \\ \nonumber 0 & -M & 0 &
-v k_{+} \\ \nonumber -v k_{+} & 0 & M & i \lambda_R \\
\nonumber 0 & -v k_{-} & -i \lambda_R & -M
\end{array}
\right] =\left[
\begin{array}{cccc}
H_1 & -T \\
-T^{\dag} & H^\dag_2
\end{array}
\right]
\end{eqnarray}
Using the similar method in Eq.(\ref{effect-K1}), we can obtain the reduced effective two-band model Hamiltonian at the vicinity of $K'$
\begin{eqnarray}
H^{\rm K'}_{\rm eff}\simeq H_{1}-T ({H^\dag_{2}})^{-1} T^{\dag} =  d_z \sigma_z-d_y\sigma_y-d_x \sigma_x\label{effect-K2}.
\end{eqnarray}

Through comparing the obtained effective Hamiltonian at $K$/$K'$ in Eqs.~(\ref{effect-K1}) and (\ref{effect-K2}) with the famous Haldane's toy model in Ref.~[\onlinecite{haldane}], we find that they share the similar characteristic form. Especially, when $k^2>\lambda^2_R/v^2$, the coefficient of $d_z$ can change its sign from positive to negative. This signals that the reduced effective Hamiltonians are definitely extended Haldane's models to exhibit a nonzero Chern number. Since the sign is directly related to $d_z$ and $d_z$ is an odd (even) function with respect to $M$ ($\lambda_R$), the sign of the resulting quantum Hall conductance should only be dependent on $M$. Moreover, the coefficient of $\sigma_z$ in Eq.~(\ref{effect-K2}) is exactly the same as that in Eq.~(\ref{effect-K1}). Therefore, according to Eq.~(\ref{TopologicalCharge}) both effective Hamiltonians at $K$ and $K'$ give rise to the same Chern number, thus the total Chern number is $\mathcal{C}=2\rm sgn(M)$.

\section{Robustness of Quantum Anomalous Hall Effect}\label{section3}
\subsection{Staggered \emph{AB} Sublattice Potential}
When graphene is doped with some magnetic atoms on the top of carbon atoms, i.e. one adatom sitting on top of the carbon atom in a $4\times4$ or $5\times5$ supercell of graphene, besides the magnetic proximity-induced exchange field and the interaction-induced Rashba spin-orbit coupling, the imbalanced \emph{AB} sublattice potentials may also be introduced.~\cite{Ding} In the following, we address the possible effect of the staggered \emph{AB} sublattice potentials on the quantum anomalous Hall state. In the low-energy limit, the effective Hamiltonian in the presence of Rashba spin-orbit coupling $\lambda_R$, exchange field $M$ and staggered sublattice potential $U$ on the basis of \{$\psi_{A\uparrow}$, $\psi_{A\downarrow}$, $\psi_{B\uparrow}$, $\psi_{B\downarrow}$\} is written as:
\begin{eqnarray}
H({\bm k})=h_0({\bm k})+h_R({\bm k})+h_M({\bm k})+h_U({\bm k}).
\end{eqnarray}
After a direct diagonalization of the above Hamiltonian, the energy dispersion can be expressed as
\begin{eqnarray}
\varepsilon({\bm k})&=&\mu\sqrt{P +  \nu \sqrt {Q}},
\end{eqnarray}
with $P$ and $Q$ being the following
\begin{eqnarray}
P&=&M^2 + U^2+\frac{1}{2} \lambda^2_R + v^2 k^2; \nonumber \\
Q&=&{\lambda^4_R/4+v^2 k^2 \lambda^2_R + 4v^2 k^2 M^2 -2 \eta M U \lambda^2_R + 4 M^2 U^2}, \nonumber
\end{eqnarray}
where $\mu=\pm1$ stands for the conduction (+) and valence (-) bands; $\nu=\pm1$ denotes the spin chirality. By imposing $k=0$, the bulk band gap $\Delta$ at valleys $K$ and $K'$ can be determined to be:
\begin{eqnarray}
\Delta=2|M-\eta U|\label{GapRelation},
\end{eqnarray}
which indicates that at valley $K$ (i.e. $\eta= + 1$), along with the increasing of the staggered \emph{AB} sublattice potential $U$ from zero, the bulk band gap $\Delta$ first decreases; at a critical $M=U$ point, the bulk gap is completely closed; when $U$ further increases to be larger than $M$, a finite bulk gap reopens, indicating a topological phase transition. However, at the other valley $K'$ (i.e. $\eta=-1$), the bulk gap $\Delta$ always increases and does not experience a topological phase transition. The different bulk band gap responses at $K$ and $K'$ are consistent with the tight-binding result discussed in Ref.~[\onlinecite{xingzhangwang}].

For a small staggered AB sublattice potential $U$, the system is in a the quantum anomalous Hall phase: both valleys induce the same unit topological charge $\mathcal{C}_{\rm K}=\mathcal{C}_{\rm K'}=1$. As long as the bulk gap is not completely closed at both $K$ and $K'$, the system should always belong to the quantum anomalous Hall phase. When $U>M$, a topological phase transition occurs at valley $K$, indicating a band inversion with Chern number becoming $\mathcal{C}_{\rm K}=-1$. Since the topology at the valley $K'$ always stays the same, i.e. $\mathcal{C}_{\rm K'}=1$, therefore the total Chern number vanishes with $\mathcal{C}=\mathcal{C}_{K}+\mathcal{C}_{K'}=0$. But the difference of Chern numbers at $K$ and $K'$ result in a quantum valley-Hall phase with valley Chern number $\mathcal{C}_v=(\mathcal{C}_K-\mathcal{C}_{K'})/2=1$. Though the resulting new phase is the same as that in a gated bilayer graphene in the presence of Rashba spin-orbit coupling and exchange field,~\cite{Tse} the major difference is that the bulk gaps at $K$ and $K'$ are simultaneously closed at some critical paramters.~\cite{qiao-preparation}

In a gated bilayer graphene, when only the Rashba spin-orbit coupling is applied, the system experiences a topological phase transition from a quantum valley-Hall phase to a two-dimensional strong topological insulator phase through tuning the gate bias between top and bottom layers.~\cite{qiao2} It is natural to hope that similar topological insulator phase can be realized by considering staggered \emph{AB} sublattice potential and Rashba spin-orbit coupling in a single layer graphene, since staggered \emph{AB} sublattice potential plays a similar role to break the out-of-plane inversion symmetry as the gate bias in bilayer graphene. However, we show that it is not the case. From Eq.~(\ref{GapRelation}), one can find that the bulk band gap is only dependent on the staggered \emph{AB} sublattice potential $U$ and the exchange field $M$ but independent of the Rashba spin-orbit coupling strength. Thus, it is obvious that the resulting bulk gap in the absence of exchange would be a constant for any Rashba spin-orbit coupling strength at a fixed staggered potential $U$. This signals that topological insulator state can not be achieved in single layer graphene through tuning Rashba spin-orbit coupling.

\subsection{Intrinsic Spin-Orbit Coupling}
Since the spin-orbit coupling (Rashba or intrinsic) in pristine graphene is very weak, one has to employ external means to enhance it. Recent ab-initio studies reported that a better approach to enlarge the Rashba spin-orbit coupling is via doping low-concentration $3d$ or $5d$ transition metal atoms on the hollow adsorption sites.~\cite{Ding,HongbinZhang} Though we only prefer the Rashba type spin-orbit coupling, the enhancement of the intrinsic one is unavoidable.~\cite{RuqianWu} In the presence of intrinsic spin-orbit coupling, Rashba spin-orbit coupling and exchange field, the continuum Hamiltonian is written as
\begin{eqnarray}
H({\bm k})=h_0({\bm k})+h_R({\bm k})+h_{SO}({\bm k})+h_M({\bm k})\label{IntrinsicH}.
\end{eqnarray}
Through diagonalizing Eq.~(\ref{IntrinsicH}) at ${\bm k}=0$, the energy spectrum at $K$ can be expressed as:
\begin{eqnarray}
\varepsilon^{K}_1&=&+M+\lambda_{SO}; \nonumber \\
\varepsilon^{K}_2&=&-M+\lambda_{SO}; \nonumber \\
\varepsilon^{K}_3&=&+\sqrt{M^2+\lambda^2_R}-\lambda_{SO}; \nonumber \\
\varepsilon^{K}_4&=&-\sqrt{M^2+\lambda^2_R}-\lambda_{SO}. \nonumber
\end{eqnarray}
And the corresponding energy spectrum at $K'$ are
\begin{eqnarray}
\varepsilon^{K'}_1&=&+\sqrt{M^2+\lambda^2_R}+\lambda_{SO}; \nonumber \\
\varepsilon^{K'}_2&=&+M-\lambda_{SO}; \nonumber \\
\varepsilon^{K'}_3&=&-\sqrt{M^2+\lambda^2_R}+\lambda_{SO}; \nonumber \\
\varepsilon^{K'}_4&=&-M-\lambda_{SO}. \nonumber
\end{eqnarray}

In general, the strength of the adatom-induced intrinsic spin-orbit coupling is an order of magnitude smaller than the induced Rashba strength, i.e. $\lambda_{SO} \ll \lambda_R$; and the exchange field is often lager than the Rashba spin-orbit coupling strength. Therefore, the resulting bulk band gaps at $K$ and $K'$ are
\begin{eqnarray}
&&\Delta_K=\varepsilon^{K}_3-\varepsilon^{K}_2=\sqrt{M^2+\lambda^2_R}+M-2\lambda_{SO}; \\
&&\Delta_{K'}=\varepsilon^{K'}_2-\varepsilon^{K'}_4=2M.
\end{eqnarray}
This indicates that the bulk gaps show different responses at $K$ and $K'$. As long as $\lambda_{SO}<(\sqrt{M^2+\lambda^2_R}+M)/2$, the quantum anomalous Hall phase would be robust against the weak intrinsic spin-orbit interaction.

From the theoretical point of view, if the intrinsic term is comparable with $\lambda_R$ and $M$ or even larger, the band gap at $K'$ changes to be the same as that at $K$, i.e.,
\begin{eqnarray}
&&\Delta_K=|\varepsilon^{K}_3-\varepsilon^{K}_2|=|\sqrt{M^2+\lambda^2_R}+M-2\lambda_{SO}|; \\
&&\Delta_{K'}=|\varepsilon^{K'}_2-\varepsilon^{K'}_3|=|\sqrt{M^2+\lambda^2_R}+M-2\lambda_{SO}|.
\end{eqnarray}
Therefore, at $\lambda_{SO}=(\sqrt{M^2+\lambda^2_R}+M)/2$, the bulk band gap completely close at both $K$ and $K'$. And it enters a new phase $\mathcal{C}=0$ when the intrinsic spin-orbit coupling $\lambda_{SO}$ further increases. Due to the presence of the exchange field $M$, the time-reversal symmetry is broken. Thus, it is no longer a intrinsic spin-orbit coupling-induced two-dimensional topological insulator. In a recent paper, it has been reported that this new phase is a time-reversal symmetry broken quantum spin-Hall phase.~\cite{YunyouYang}

\section{Theory of Metal Adsorption on $3\times3$ and $4\times4$ supercells of graphene}\label{section4}
In the previous sections, we have assumed that all involved parameters are uniformly distributed on each atomic site of the graphene sheet. However, in a more realistic graphene sample, the atom dopants are usually adsorbed on graphene with a low concentration to avoid the direct transport through dopants themselves. For example, the mostly adopted systems in the ab-initio study are $3\times3$, $4\times4$, $5\times5$ and $7\times7$ supercells of graphene. For $3\times3$ supercells, valleys $K$ and $K'$ are coupled to the $\Gamma$ point resulting in the mixtures of valleys, but in the last three kinds of supercells valleys $K$ and $K'$ are separated and well-defined to be good quantum numbers. Therefore, in the following discussion we only consider two representing $3\times3$ and $4\times4$ supercells of graphene using the tight-binding methods.
\begin{figure}
\includegraphics[width=9cm,angle=0]{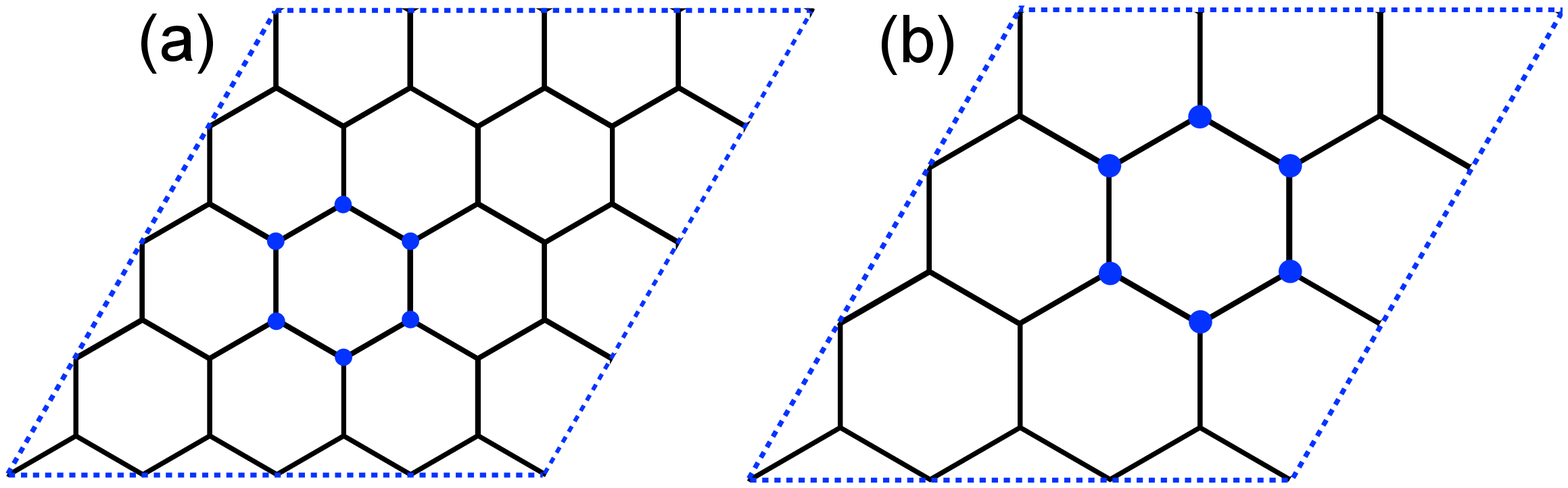}
\caption{Schematic plot of (a) $4\times4$ and (b) $3\times3$ supercells of graphene. On-site crystal field potential, Rashba spin-orbit coupling, and exchange field are only considered on the highlighted sites.} \label{non-uniform-setup}
\end{figure}

There are three highly possible adsorption points in graphene: top, bridge, and hollow.~\cite{Ding,Cohen} In Ref.~[\onlinecite{Ding}], we have shown that only the hollow-site adsorption can open a nontrivial bulk gap to achieve the quantum anomalous Hall state. An obvious characteristic of this kind of adsorption is that the induced effects are non-uniformly distributed in the supercell, i.e. the six nearest carbon atoms under the metal adatom experience the largest exchange field $M$ and Rashba spin-orbit coupling $t_R$, while for the other carbon atoms the longer the distance from the adsorption site, the smaller the induced interactions. Another most important term arisen from the adsorption is the crystal field stabilization energy $V_0$ (also known as on-site energy), which is the main factor coupling valleys $K$ and $K'$ in the $3\times3$ supercell of graphene when Rashba and exchange effects are absent.

As shown in Fig.~\ref{non-uniform-setup}, we schematically plot the $4\times4$ (a) and $3\times3$ (b) supercells of graphene. To emphasize the inhomogeneity, we only consider the externally induced effects ($t_R$, $M$, and $V_0$) on the six highlighted atomic sites of the supercells, while the remaining atomic sites are modeled as a pristine graphene. In our simulation, the effective tight-binding Hamiltonian is the same as Eq.~(\ref{SGHamiltonian}) by setting $t_{SO}=0$.

\subsection{$3\times3$ supercell of graphene}
\begin{figure}
\includegraphics[width=8cm,angle=0]{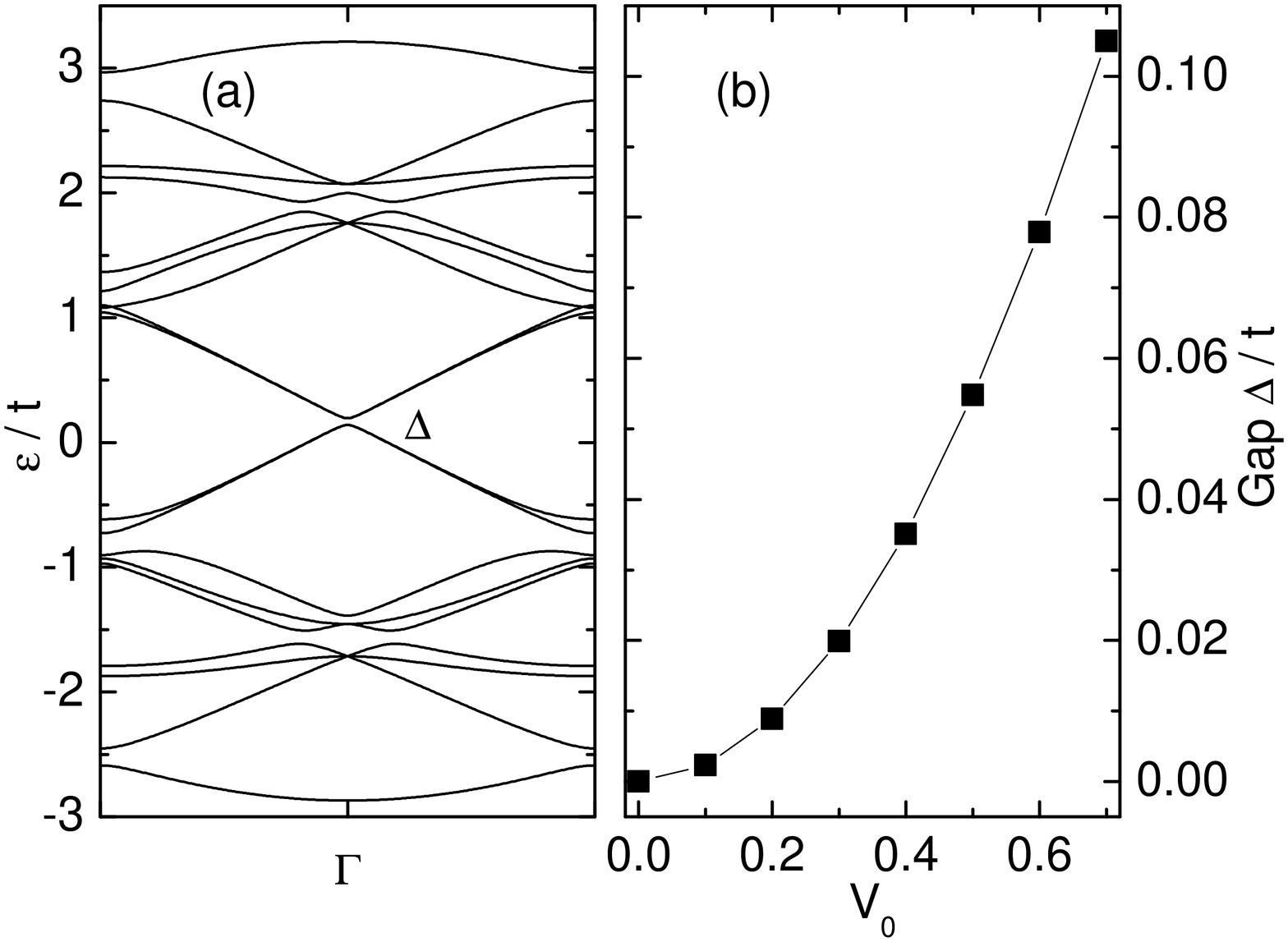}
\caption{(a) Bulk band structure of the $3\times3$ supercell of graphene in the presence of only on-site potential $V_0/t=0.50$. A band gap $\Delta$ opens at $\Gamma$ point. (b) Bulk gap $\Delta$ quadratically increases as a function of $V_0$.} \label{OnSitePotentialDependence}
\end{figure}
Figure~\ref{OnSitePotentialDependence} plot the bulk band structure of the $3\times3$ supercell of graphene in the presence of only on-site potential and the resulting gap dependence as a function of the on-site potential. One can observe that a trivial band gap $\Delta$ opens at $\Gamma$ point [see Fig.~\ref{OnSitePotentialDependence}(a)]; and the opened band gap increases quadratically as a function of the on-site potential strength $V_0$. This confirms that it is the adsorption-induced on-site potential that couples valleys K and K' to open a band gap, consistent with the ab-initio calculation result in Fig.~6 of Ref.~[\onlinecite{Ding}].

\begin{figure}
\includegraphics[width=8cm,angle=0]{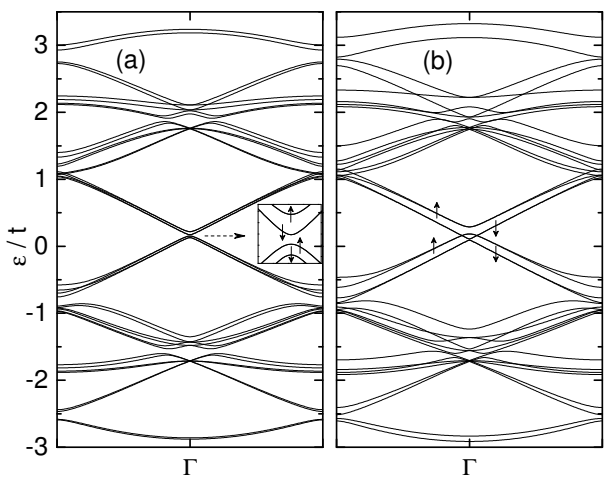}
\caption{Bulk band structure of the $3\times3$ supercell of graphene in the presence of on-site potential $V_0/t=0.50$ and exchange field $M$. (a) $M/t=0.05$; (b) $M/t=0.20$. $\uparrow$ and $\downarrow$ denote up and down spin polarization.} \label{band-33M}
\end{figure}

When the exchange field $M$ is further included, the bulk gap first decreases due to the relative shift between the spin-up polarized valence band and the spin-down polarized conduction band [see Fig.~\ref{band-33M}(a)]. For even larger $M$ as shown in Fig.~\ref{band-33M}(b), one can observe that the gap is completely closed and the bands with opposite spin polarization cross.

\begin{figure}
\includegraphics[width=8cm,angle=0]{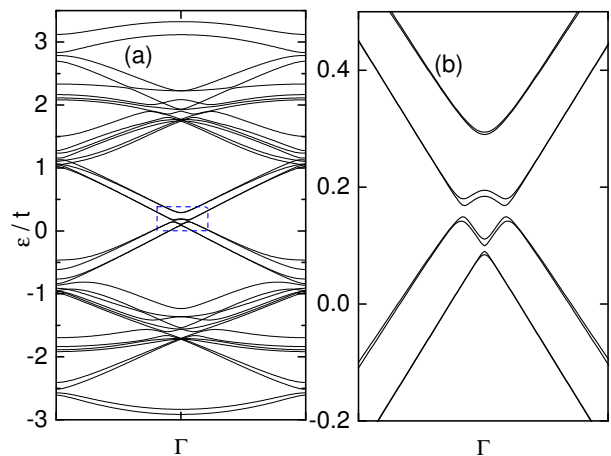}
\caption{(a) Bulk band structure of the $3\times3$ supercell of graphene in the presence of on-site potential $V_0/t=0.50$, exchange field $M/2=0.20$, and Rashba spin-orbit coupling $t_R/t=0.05$. (b) The magnification of the band gap selected by dashed square.} \label{33RSOM}
\end{figure}

If Rashba spin-orbit coupling is considered in addition to the exchange field, we find that for small exchange field since the original band gap from on-site potential does not close, and the not-so-large Rashba spin-orbit coupling can only further reduce the band gap. However, for large exchange field, the situation becomes completely different, i.e. the Rashba spin-orbit coupling opens a new band gap at the band-crossing points as shown in Fig.~\ref{33RSOM}.

\begin{figure}
\includegraphics[width=6cm,angle=0]{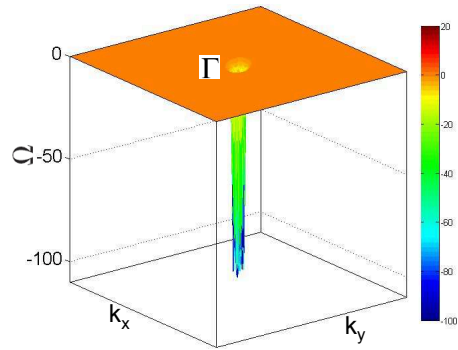}
\caption{Total Berry curvature distribution $\Omega$ in the momentum space of the occupied valence bands below the band gap. Only those around $\Gamma$ point nonzero and share the same negative sign.} \label{BerryCurvature33}
\end{figure}

To explore the nontrivial topology of the newly formed insulating phase, we plot the total Berry curvature distribution $\Omega(k_x,k_y)$ of the occupied valence bands below the gap in Fig.~\ref{BerryCurvature33}. One can find that the nonzero Berry curvatures are mainly located around $\Gamma$ point and share the same negative sign, suggesting a nonzero Chern number. Through an integration of Berry curvatures over the first Brillouin zone, the Chern number is calculated to be $\mathcal{C}=2$.

Till now, we can conclude that in the $3\times3$ supercells of graphene, the prerequisite to realize the quantum anomalous Hall effect is that the exchange field should be large enough to close the trivial band gap arisen from the crystal field stabilization energy $V_0$. To our surprise, another separate work~\cite{JiangHua} proves that the randomness of the adsorption sites can exponentially diminish the trivial band gap determined by crystal field stabilization energy. This information manifests the high possibility of engineering the long-sought quantum anomalous Hall state in graphene through adsorbing random magnetic atoms.
\subsection{$4\times4$ supercell of graphene}
\begin{figure}
\includegraphics[width=8cm,angle=0]{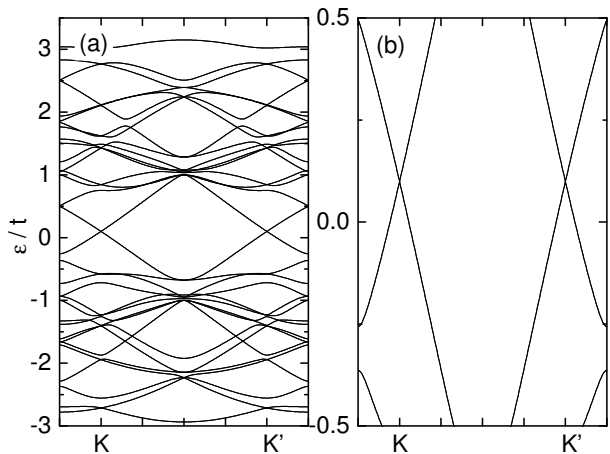}
\caption{(a) Bulk band structure of the $4\times4$ supercell of graphene in the presence of on-site potential $V_0/t=0.50$. (b) Magnification of bands near the Dirac crossing point. No gap opens at $K$ and $K'$.} \label{U-44}
\end{figure}

Let us now study the $4\times4$ supercell of graphene case. Figure~\ref{U-44}(a) plots the whole bulk band structure of the $4\times4$ supercell of graphene in the presence of only on-site potential $V_0$, and Figure~\ref{U-44}(b) magnifies the bands at the low-energy region. One can observe that the bands at $K$ and $K'$ exhibit linear Dirac-type dispersion without opening a band gap, which are completely different from the result of $3\times3$ supercell of graphene. Comparing with the band structure of the pristine graphene, one can find that the bands at the low-energy regime are similar except a Fermi-level shifting.

\begin{figure}
\includegraphics[width=8cm,angle=0]{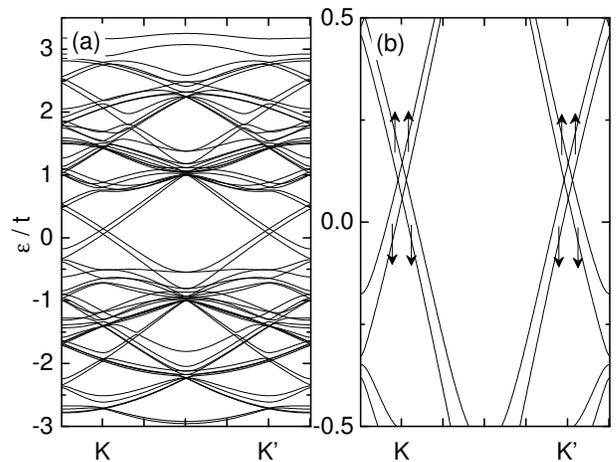}
\caption{(a) Bulk band structure of the $4\times4$ supercell of graphene in the presence of on-site potential $V_0/t=0.50$ and exchange field $M/t=0.20$. (b) Magnification of bands near the Dirac crossing point. Arrows are used to denote the up and down spin polarizations.} \label{U-M44}
\end{figure}

As shown in Fig.~\ref{U-M44}, when the exchange field is further considered, the doubly-degenerate bands become spin-split with spin-up bands up-ward shifting and spin-down bands down-ward shifting. This resembles the spin-split of graphene in the presence of uniformly distributed exchange field.~\cite{qiao}

\begin{figure}
\includegraphics[width=8cm,angle=0]{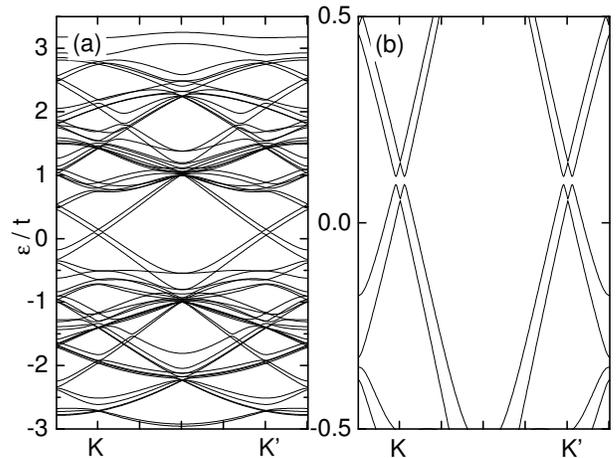}
\caption{(a) Bulk band structure of the $4\times4$ supercell of graphene in the presence of on-site potential $V_0/t=0.50$, exchange field $M/t=0.20$, and Rashba spin-orbit coupling $t_R/2=0.05$. (b) Magnification of bands near the Dirac crossing point. Bulk band gaps open at $K$ and $K'$} \label{RSO-U-M44}
\end{figure}

Figure~\ref{RSO-U-M44} plots the bulk band structure of the $4\times4$ supercell of graphene in the presence of on-site energies, exchange field, and
Rashba spin-orbit coupling. One can observe that the Rashba spin-orbit coupling term opens a gap at the spin-up and spin-down band-crossing points near valleys $K$ and $K'$. This gap formation should be nearly the same as that pointed out in Ref.~[\onlinecite{qiao}]. Therefore, the corresponding Chern number should be $\mathcal{C}=2$. These tight-binding band structures reproduces the ab-initio band structures demonstrated in Ref.~[\onlinecite{Ding}]

\section{Conclusions} \label{section5}
In this paper, we discuss the physical origins of the formation of quantum anomalous Hall effect in graphene due to the presence of Rashba spin-orbit coupling $\lambda_R$ and exchange field $M$ using a continuum model. We show that in the limit of $M/\lambda_R\gg1$, the quantization of the Hall conductance arises from Skyrmions carried by the real spin textures and Merons carried by \emph{AB} sublattice pseudo-spin textures at $K$ and $K'$; in the other limit $\lambda_R/M \gg1$, the four-band low-energy Hamiltonian is reduced to an extended Haldane's model, giving rise to a nonzero Chern number $\mathcal{C}=1$ at either $K$ or $K'$.

We demonstrate that the quantum anomalous Hall phase is robust against weak staggered \emph{AB} sublattice potential $U$ or intrinsic spin-orbit coupling $\lambda_{SO}$. In the presence of a moderate staggered AB sublattice potential, the system undergoes a phase transition from a quantum anomalous Hall phase to a quantum valley-Hall phase if $U>M$. Alternatively, when a larger intrinsic spin-orbit coupling is applied, graphene in a quantum anomalous Hall phase transitions to a time-reversal-symmetry broken quantum spin-Hall phase~\cite{YunyouYang} at $\lambda_{SO}=(\sqrt{M^2+\lambda^2_R}+M)/2$.

Using a tight-binding model Hamiltonian, we reproduce all the ab-initio band structures~\cite{Ding} (at the low-energy level) of doping magnetic atoms on the hollow site of the $3\times3$ and $4\times4$ supercells of graphene by considering the on-site energy (crystal field stabilization energy), exchange field and Rashba spin-orbit coupling on only a circle of six atomic sites, and explain the formations of the quantum anomalous Hall state in the $3\times3$ and $4\times4$ supercells of graphene. For the $3\times3$ supercell of graphene, we show that the crystal field stabilization energy is crucial to couple valleys $K$ and $K'$ to open a trivial bulk band gap at $\Gamma$ point in the absence of exchange field and Rashba spin-orbit coupling. We also find that only when the exchange field is large enough to close the trivial band gap from the crystal field stabilization energy, a nontrivial bulk band gap exhibiting the quantum anomalous Hall effect can be opened due to the presence of Rashba spin-orbit coupling. For the $4\times4$ supercell of graphene, due to the separation of valleys, no band gap opens when only the crystal field stabilization energy is present. When exchange field and Rashba spin-orbit coupling are considered simultaneously, the physical mechanism to open a bulk gap is exactly the same as that in the presence of uniformly distributed parameters.~\cite{qiao}

\section{Acknowledgements}
Z.Q. was supported by the NSF (Grant No.~DMR 0906025) and the Welch Foundation (Grant No.~F-1255). Q.N. was supported by the DOE (Grant No.~DE-FG03-02ER45958,
Division of Materials Science and Engineering) and the Texas Advanced Research Program. H.J. was supported by the CPSF (Grant No.~20100480147 and No.~201104030). Y.Y. was supported by the NSF of China (Grants No.~10974231 and No.~11174337) and the MOST Project of China (Grant No.~2011CBA00100).

\end{document}